# Cosmological Constant and Statistical Lensing of Giant Arcs

Xiang-Ping Wu[1,2] and Shude Mao[1,3]

[1] Beijing Astronomical Observatory, Chinese Academy of Sciences, Beijing 100080, China

[3] MS 51, Center for Astrophysics, 60 Garden Street, Cambridge, MA 02138

---

[2]Present address: Department of Physics, University of Arizona, Tucson, AZ 85721

## ABSTRACT

Using a singular isothermal sphere model for the matter distribution of foreground clusters of galaxies, we study the statistics of giant arcs in flat cosmologies with and without a cosmological constant. We find that the relative number of arcs predicted within $z = 1$ in a universe with $\Omega_0 = 0.3$ and $\lambda_0 \equiv \Lambda/(3H_0^2) = 0.7$ is a factor of $\sim 2$ larger than the one in the Einstein-de Sitter universe ($\Omega_0 = 1, \lambda_0 = 0$). For a luminosity-dependent evolution model of the number density of background galaxies that accounts for the over-density of faint blue galaxies at $z_s \approx 0.4$, the Einstein-de Sitter cosmological model predicts that about 5% of clusters of galaxies with an X-ray luminosity $L_x > 2 \times 10^{44}$ erg s$^{-1}$ should have giant arcs with length-to-width ratio larger than 10. This is a factor of $\sim 4$ lower than the observed fraction in the gravitational lensing survey of distant X-ray selected EMSS clusters of galaxies, indicating that the matter distribution of clusters of galaxies deviates significantly from simple isothermal spheres or/and the presence of a significant cosmological constant. It is profitable to further study the constraint on the cosmological constant from giant arcs using more realistic cluster models.

*Subject headings:* cosmology: theory — galaxies: clusters: general — – gravitational lensing

## 1. Introduction

The cosmological constant has been invoked many times over history (Caroll, Press & Turner 1992, and references therein; Efstathiou 1995). In particular, the recent measurements of the Hubble constant from Cepheid variables of nearby galaxies (Pierce et al. 1994; Freedman et al. 1994; Tanvir et al. 1995) yield an age of the universe of $8 - 13$ Gyr, while the oldest stars in globular clusters of our Galaxy are estimated to be about $15.8 \pm 2.1$ Gyr (Bolte & Hogan 1995). It seems that a nonzero cosmological constant is the only way at present out of the conflict (Ostriker & Steinhardt 1995). Turner (1990) and Fukugita, Futamase & Kasai (1990) first explicitly pointed out that the introduction of a large cosmological constant significantly changes the number of multiply imaged quasars (see also Paczyńksi & Gorski 1981; Alcock & Anderson 1986; Gott et al. 1989). Thus, gravitational lenses can be used to set robust constraint on the cosmological constant. Recent studies find that $\lambda_0 \equiv \Lambda/(3H_0^2)$ cannot be larger than 0.7 (Fukugita et. al. 1992; Kochanek 1992, 1993, 1995). This conclusion is little changed by reasonably expected galaxy evolution (Mao & Kochanek 1994; Rix et al. 1994).

In this paper, following the logic of the above studies, we consider the influence of the cosmological constant on the prediction of giant arcs. Giant arcs are the images of background galaxies gravitationally distorted by foreground clusters of galaxies. Compared with multiply imaged quasars, giant arcs have two major differences. The first is that these systems probe the mass distribution on the cluster scale. The second is that the source population is not quasars with known redshifts but the distant faint galaxies ($z \approx 1$), which are still largely unexplored. This brings some uncertainty to the study of arcs, but also provides an exciting opportunity to probe the evolution of the background galaxies.

At the time of this writing, more than 30 giant arcs are known (Surdej & Soucail 1993; Fort & Mellier 1994; Wu 1995). Some of these were discovered serendipitously, which makes statistical comparison rather difficult. Fortunately, a well-defined sample of the gravitationally lensed arc survey in the 41 EMSS distant clusters have recently become available (Gioia & Luppino 1994; Hammer 1995; Luppino et al. 1995). This makes a comparison between the theoretical predictions and observations a timely necessity. The present work intends to fulfill this task. Furthermore we want to examine whether a cosmological constant significantly changes the theoretically predicted number of giant arcs, their angular positions and whether the observations provide any constraints on its presence. To fulfill these goals, we adopt a simple analytical formalism (cf. Wu & Hammer 1993) and present the results in §2. In §3 we compare the results with the observations and discuss the uncertainties and implications of our results. Throughout the paper, we adopt a Hubble constant of $H_0 = 50$ km s$^{-1}$ Mpc$^{-1}$.

## 2. Method and Results

To study the statistical properties of giant arcs, we need to know the spatial and dynamical properties of foreground clusters and the number and luminosity distributions of distant background galaxies. Note that the statistical properties of clusters and background galaxies are often derived using a specific cosmology. These properties need to be recomputed when the cosmological constant is involved.

Clusters of galaxies have complex structures. Most theoretical studies in the past (Grossman & Narayan 1988, 1989; Nemiroff & Dekel 1989; Wu & Hammer 1993; Miralda-Escudé 1993a, b; Bergmann & Petrosian 1993; Grossman & Saha 1994) used simple analytical potential forms for the clusters. However, the recent studies by Bartelmann & Weiss (1994) and Bartelmann, Steinmetz, & Weiss (1995) reached another level of sophistication: they used numerically simulated clusters to study their capabilities of forming arcs. In this paper, for simplicity, we adopt the singular isothermal sphere (SIS) model for the matter distribution of galaxy clusters. The analytical SIS model is adopted because our emphasis is on the comparison between an Einstein-de Sitter cosmology ($\Omega_0 = 1, \lambda_0 = 0$, hereafter standard cosmology) and nonzero $\lambda_0$ cosmologies. Furthermore, the SIS model gives lensing probabilities that are (30-40)% off from the more sophisticated model by Bartelmann et al. (1995, Fig. 9, see however Bergmann & Petrosian 1993). In the SIS model, a cluster is characterized solely by its velocity dispersion, $\sigma_v$. As the distant, X-ray selected EMSS clusters of galaxies used in the gravitationally lensed arc survey (Gioia & Luppino 1994) show strong evolution in the X-ray luminosity ($L_x$) with cosmic epoch, Henry et al. (1992) divided their luminosity function into three redshift shells ranging from 0.14 to 0.6. Each shell is represented by a power law

$$\eta(L_{x,44}) = K L_{x,44}^{-\alpha}, \tag{1}$$

where $L_{x,44}$ is the X-ray luminosity in units of $10^{44}$ erg s$^{-1}$, and $K$ and $\alpha$ take different values at the three redshift shells. The Henry et al. (1992) luminosity function was derived using the standard cosmology. How does $\lambda_0$ change the luminosity function? In general, the introduction of $\lambda_0$ increases the comoving distance and volume for a given redshift range and therefore, reduces the number *density*. The introduction of $\lambda_0$ also increases the estimate of the intrinsic luminosity corresponding to a given flux. However, in principle it does not alter the number counts of a population. In practice, the situation is slightly more complicated. For example, the $1/V_a$ method was used to derive the X-ray luminosity function of galaxy clusters (Henry et al. 1992). Cosmology was then introduced in the process of correcting for different sensitivities in various parts of the sky. Therefore, the X-ray luminosity function of galaxy clusters has to be recomputed for different cosmologies. We find, however, that $\lambda_0$ does not significantly change the luminosity function since $\lambda_0$ reduces the number density of clusters for a given flux and meanwhile increases the estimate of the intrinsic luminosity. As a result of these two effects, the luminosity function remains roughly unchanged.

To convert the X-ray luminosity function into the distribution of velocity dispersion, we employ the *quasi*-Faber-Jackson relation for the X-ray selected clusters of galaxies (Quintana & Melnick 1982; see also Wu & Hammer 1993):

$$L_x = 10^{32.72}\sigma_v^{3.94} \text{ erg s}^{-1},\qquad(2)$$

where $\sigma_v$ is in units of km s$^{-1}$. The more recent X-ray observations have shown that the relationship between the cluster X-ray luminosity and temperature is $L_x = 1.1 \times 10^{43} T^2$ erg s$^{-1}$ (Bahcall & Cen 1994, and references therein), where $T$ is expressed in units of keV. Optical observations of the cluster galaxies and X-ray observations of the hot cluster gas give the best fit of $\sigma_v$ versus $T$: $\sigma_v = 400\, T^{1/2}$ km s$^{-1}$ (Lubin & Bahcall 1993). Combining these two relations yields

$$L_x = 10^{32.64}\sigma_v^4 \text{ erg s}^{-1},\qquad(3)$$

in good agreement with eq. (2).

We consider spiral galaxies as the background sources and assume a constant magnification over the luminous disks of the sources. Both no-evolution and evolution models of the number density of galaxies will be considered. For the no-evolution model, we take the normalization of the local spiral galaxies with a Schechter luminosity function:

$$\phi(L_g, 0)dL_g = \phi^*\,(L_g/L_g^*)^\alpha\,\exp(-L_g/L_g^*)\,dL_g/L_g^*,\qquad(4)$$

where $\phi^* = 1.95 \times 10^{-3}$ Mpc$^{-3}$, $\alpha = -1.1$ and $M^* = -21.4$ (Efstathiou, Ellis, & Peterson 1988). For the evolution model, we employ the empirical form of luminosity-dependent evolution given by Broadhurst, Ellis & Shanks (1988)

$$\log\phi(L_g, z_s) = \log\phi(L_g, 0) + (0.1z_s + 0.2z_s^2)\log\left[\frac{\phi(L_g,0)}{\phi(L_{g,max},0)}\right],\qquad(5)$$

in which $L_{g,max}$ is the truncated luminosity at $M_{max} = -23.5$. The galaxy redshift should be restricted within $z_s < 1$ due to the failures of this evolutionary scenario and of the

$k$−correction that reads

$$k(z_s) = -0.05 + 2.3z_s + 2.55z_s^2 - 4.89z_s^3 + 1.85z_s^4 \qquad (6)$$

for spirals (King & Ellis 1985). Note that the introduction of $\lambda_0$ does not change the number counts of background galaxies versus their apparent magnitude and redshift, but it does increase the intrinsic absolute luminosity corresponding to a given apparent magnitude. $\lambda_0$ does not change the $k$-term either as it only reflects the spectral shift due to the redshift term.

The total probability $P(z_s, \mu)$ of a background point source at redshift $z_s$ magnified by a factor of greater than $\mu$ by the foreground intervening clusters at the redshift shell of $(z_{d1}, z_{d2})$ is then

$$P(z_s, \mu) = \frac{16\pi^3}{(\mu-1)^2} \int_{z_{d1}}^{z_{d2}} (1+z_d)^3 \left(\frac{D_d D_{ds}}{D_s}\right)^2 dr_{prop,z_d} \cdot \left(\int_{L_x}^{\infty} \left(\frac{\sigma_v}{c}\right)^4 \eta(L_{x,44}) dL_{x,44}\right), \qquad (7)$$

where $D_d$, $D_s$ and $D_{ds}$ are the angular diameter distances to the foreground galaxy cluster, to the background source and from the cluster to the source, respectively, and $dr_{prop,z_d}$ is the proper distance around redshift $z_d$.

We only investigate flat cosmologies, i.e., cosmologies with $\Omega_0 + \lambda_0 = 1$. The proper distance is given by (Fukugita et al. 1992)

$$dr_{prop,z} = \frac{c}{H_0} \frac{dz}{(1+z)\sqrt{\Omega_0(1+z)^3 + 1 - \Omega_0}}. \qquad (8)$$

The angular diameter distance from redshift $z_1$ to $z_2$ is

$$d(z_1, z_2) = \frac{c}{H_0} \frac{1}{1+z_2} \int_{z_1}^{z_2} \frac{dz}{\sqrt{\Omega_0(1+z)^3 + 1 - \Omega_0}} \qquad (9)$$

Clearly, $D_d = d(0, z_d)$, $D_s = d(0, z_s)$ and $D_{ds} = d(z_d, z_s)$.

Giant arcs are defined as the gravitationally elongated images with axis ratios $L/W \geq 10$ (Wu & Hammer 1993), here $L$ and $W$ are the length and width of the images, respectively. Because of the identity of magnification $\mu$ with axial ratio $L/W$ in the SIS model, magnification probability $P(z_s, \mu)$ is the same as the probability of a background source being elongated by a factor of greater than $L/W$: $P(z_s, L/W)$. Convolving the number density of spiral galaxies with the lensing probability of clusters of galaxies and integrating over redshift range $(z_d, z_s)$ give the total number of giant arcs one would expect to observe above the limiting magnitude $B$:

$$N(B, L/W, L_x) = 4\pi \left(\frac{c}{H_0}\right)^3 \int_{z_d}^{z_s} P(z_s, L/W)(1+z_s)^3 D_s^2 dr_{prop,z_s} \int_{L_{g,min}/L_g^*}^{L_{g,max}/L_g^*} \phi(L_g, z_s) dL_g, \tag{10}$$

where $L_{g,min}/L_g^* = min\{0.1, (L/W)^{-1} 10^{-0.4(B-B^*)}\}$, $B^* = M^* + 5\log(D_{L,z_s}/10\mathrm{pc}) + k(z_s)$ and $D_{L,z_s}$ is the luminosity distance of the source at $z_s$. Here the luminosity function of spiral galaxies has also been truncated at the faint end at $L_g = 0.1 L_g^*$ since the arc sources are primarily composed of those moderately bright galaxies having $0.1 L_g^* < L_g < L_g^*$ (Wu & Hammer 1993). Our results are insensitive to the cutoff.

We perform the above numerical computations by choosing the same parameters as those used in the gravitationally lensed arc survey in a complete subsample of rich clusters of galaxies, selected from the EMSS catalog (Le Fèvre et al 1994; Gioia & Luppino 1994; Hammer 1995; Luppino et al. 1995): $0.15 \leq z_d \leq 0.6$, $B \leq 23.5$ and $L_x \geq 2 \times 10^{44}$ erg s$^{-1}$ or $L_x \geq 4 \times 10^{44}$ erg s$^{-1}$. The total number of the predicted giant arcs ($L/W \geq 10$) and the frequency of arc detections, defined as the ratio of the total arcs to the surveyed clusters, are shown in Table 1 for different cosmological models. The total clusters within $0.15 \leq z_d \leq 0.6$ are estimated to be 7500 for $L_x \geq 2 \times 10^{44}$ erg s$^{-1}$ and 1800 for $L_x \geq 4 \times 10^{44}$ erg s$^{-1}$. Note again that in the above calculations the number counts (rather than densities) of the X-ray selected clusters of galaxies and of background galaxies are conserved for a given redshift range in different cosmological models. It turns out that the expected number of giant arcs depends sensitively on the adopted $\lambda_0$. The frequency of arc detections for the $\Omega_0 = 0, \lambda_0 = 1$ cosmology is larger than the standard cosmology by a factor of $\sim 3$. In a less $\lambda_0$ dominated universe with $\Omega_0 = 0.3$ and $\lambda_0 = 0.7$, the expected giant arcs are $\sim 2$ times more than those in the standard cosmology. This result is in good agreement with the result given by eq. (33) in Caroll et al. (1992) if one assumes a constant comoving number density of lenses and a source redshift close to $\sim 1$. It is encouraging to see that although the lenses (clusters) and sources (galaxies) are at relatively lower redshifts than those in multiply imaged systems, giant arcs can still be used as a probe for the existence of the cosmological constant.

The strong evolutionary scenario of number and luminosity of galaxies gives rise to 7 – 8 times more galaxies to $z_s = 1$ than the no-evolution model. As a result, the predicted arcs and detection fraction are much larger in the evolution model. However, the relative ratio of the arc number in a nonzero cosmological universe to the one in the standard model remains the same for the two models of galaxy spatial distribution.

To test the dependence of arc separations from the centers of their associated clusters of galaxies on the cosmological constant $\lambda_0$, we plot in Figure 1 the predicted giant arcs versus their positions from the cluster centers for a cosmology with $\Omega_0 = 0.3$ and $\lambda_0 = 0.7$ and for the standard cosmology, respectively. It appears that the difference of the distributions of arc positions between these two models are rather insignificant in shape, though the distribution in a $\lambda_0$ dominated universe peaks towards a slightly large distance from the cluster centers. Therefore, the arc position is not an efficient indicator of $\lambda_0$, similar to the case in the multiply imaged systems (e.g., Kochanek 1992).

In Figure 2 we have also illustrated the influence of $\lambda_0$ on the axis ratios $L/W$. The $\lambda_0$ dominated cosmological model provides a relatively smaller ratio of arcs with $L/W \geq 4$ to giant arcs ($L/W \geq 10$) than the matter dominated cosmological model. The dependence of arc ratio on the cosmology is introduced by the lower limit $L_{g,min}/L_*$ in eq. (10). This lower limit is determined by both $L/W$ and $D_L$. As the cosmological constant increases $D_L$, therefore compared with the standard cosmology, arcs need to have higher magnification (i.e., higher axis ratio) to be detected in a flux limited survey. However, it seems unlikely that a $\lambda_0 \approx 1$ universe can actually resolve the puzzle of too few medium arcs seen in the arc survey (Hammer et al. 1993; Hammer 1995).

## 3. Discussion

The gravitational lensing arc survey (Le Fèvre et al 1994; Gioia & Luppino 1994; Hammer 1995; Luppino et al. 1995) found 9 arcs out of 39 clusters with $L_x > 2 \times 10^{44}$ erg/s, and 7 arcs out of 23 clusters with $L_x > 4 \times 10^{44}$ erg/s. Although the statistics are still somewhat limited, it is however clear that the fraction of X-ray luminous clusters with giant arcs are rather high. For the SIS model of clusters of galaxies, the simple no-evolution model for distant spirals fails by at least a factor of 10, whatever the cosmology would be. Even with the evolution model derived from the faint blue galaxies (Broadhurst, Ellis, & Shanks 1988), the observed fraction is still a factor of $\sim 3 - 5$ times higher than the prediction of the standard cosmology. The $\Omega_0 = 0.3, \lambda_0 = 0.7$ model is a factor of 2 - 3 lower than the observation. But, an extreme $\lambda_0$ dominated universe can certainly remove this discrepancy (Table 1). We now discuss other possibilities to solve the discrepancy.

For the standard cosmological model, we can still reconcile the theoretical prediction with the observation by further increasing the number counts of galaxies at high redshift ($z_s \sim 1$) by a factor of $3 - 5$ (see Table 1). Note this increase is on top of the evolution model described by eq. (5), which already predicts a 3-4 times overdensity of faint blue galaxies seen at intermediate redshift $z_s \approx 0.4$, and a factor of 7-8 overdensity at redshift of 1 (cf. Koo & Kron 1992). There is so far no observational evidence to support such rapid evolution around $z = 1$. Another related possibility is that the arc galaxies observed so far are located mostly beyond $z_s = 1$, the source cutoff redshift used in the evolution model of galaxies. This is however inconsistent with most identified arc redshifts ($z_s \approx 0.7 - 0.8$). This possibility can be further tested when the redshifts of all the giant arcs are measured.

Our theoretical prediction is based on the simple SIS model for the clusters of galaxies. The SIS model predicts the presence of counter arcs that are rarely observed. The lack of such features indicates that the model used here is too simplistic. The discrepancy may therefore have arisen from the naive modeling of matter distribution in the clusters of galaxies. Indeed, statistical lensing of giant arcs based on more realistic models can increase the predicted number of giant arcs and change the axis ratio distribution. Bergmann & Petrosian (1993) examined a softened isothermal sphere with a finite core radius and an elliptical potential for cluster of galaxies. They found an excess of giant arcs as compared with a simple SIS. In particular, they pointed out that the number of small magnification events with respect to the highly magnified images can be greatly reduced if the source extension is involved. In fact, even for spherical density profiles, the statistical properties can be very different (Wu & Hammer 1993; Grossman & Saha 1994), while asymmetrical matter distributions are usually needed in the modeling of the arc-cluster systems (Fort & Mellier 1994). However, it remains to be seen whether a more realistic model for clusters can explain the discrepancy and a firmer limit on $\lambda_0$ can be set.

The errors introduced from the luminosity-velocity dispersion relation for clusters of galaxies can be estimated using the recent empirical formula of eq. (3) instead eq. (2). It turns out that the predicted arcs are reduced by 20% irrespective of the cosmological and evolutionary models. Therefore, this does not change our main results.

To summarize, the introduction of the cosmological constant can increase the theoretically predicted number of giant arcs, providing a possible way out of the discrepancies between the small expected number of giant arcs in standard model and the large observed fraction of giant arcs in the arc survey of the subsample of EMSS clusters.

If there is a nonzero cosmological constant, as were suggested by the recent observations (see summary in Ostriker & Steinhardt 1995), the fractions of rich clusters with giant arcs will be increased relative to the standard cosmology. The increase should be more or less equally important for arcs with various axial ratios. This offers a potential way to distangle the effect of a cosmological constant and the deviations of potentials of clusters of galaxies from SIS, as the latter can dramatically alter the distributions of axial ratios (Bergmann & Petrosian 1993). It seems difficult at present to set a robust constraint on the cosmological constant from giant arcs. However, this will become feasible in the future when the dynamical properties of clusters of galaxies and the evolution of galaxies are better understood.

SM wishes to thank the hospitality of the Beijing Astronomical Observatory, in particular Profs. Jian-Sheng Chen and Qibin Li. We also acknowledge the constructive comments by an anonymous referee. This work was supported by the National Science Foundation of China and a CfA postdoctoral fellowship.

Table 1    Predicted number and fraction of giant arcs

| $\Omega_0$ | $L_x \geq 2 \times 10^{44}$ erg s$^{-1}$ | | | | $L_x \geq 4 \times 10^{44}$ erg s$^{-1}$ | | | |
|---|---|---|---|---|---|---|---|---|
| | no-evolution | | evolution | | no-evolution | | evolution | |
| | arcs | fraction | arcs | fraction | arcs | fraction | arcs | fractions |
| 0.0 | 157 | 2.1 | 1060 | 14 | 89 | 4.9 | 590 | 33 |
| 0.1 | 121 | 1.6 | 844 | 11 | 69 | 3.8 | 470 | 26 |
| 0.2 | 100 | 1.3 | 713 | 9.5 | 57 | 3.1 | 397 | 22 |
| 0.3 | 85 | 1.1 | 624 | 8.3 | 48 | 2.7 | 347 | 19 |
| 0.4 | 75 | 1.0 | 558 | 7.4 | 42 | 2.4 | 310 | 17 |
| 0.5 | 66 | 0.9 | 507 | 6.8 | 38 | 2.1 | 281 | 16 |
| 0.6 | 60 | 0.8 | 467 | 6.2 | 34 | 1.9 | 258 | 14 |
| 0.7 | 55 | 0.7 | 433 | 5.8 | 31 | 1.7 | 239 | 13 |
| 0.8 | 50 | 0.7 | 404 | 5.4 | 29 | 1.6 | 223 | 12 |
| 0.9 | 47 | 0.6 | 380 | 5.1 | 26 | 1.5 | 209 | 12 |
| 1.0 | 44 | 0.6 | 359 | 4.8 | 25 | 1.4 | 197 | 11 |

---

Fig. 1.— The number of giant arcs predicted over the whole sky versus angular separation. We adopt a luminosity-dependent evolution model for the galaxy number density to $z_s = 1$ and search for arcs behind clusters with an X-ray luminosity in excess of $2 \times 10^{44}$ erg s$^{-1}$. The arcs all have a length-to-width ratio larger than 10 ($L/W > 10$). The solid histogram is for $\Omega_0 = 1, \lambda_0 = 0$ universe, and the dashed histogram is for $\Omega_0 = 0.3, \lambda_0 = 0.7$.

Fig. 2.— The ratio of the arcs with $L/W > 4$ to the giant arcs with $L/W > 10$ as a function of the matter density $\Omega_0$ ($\Omega_0 + \lambda_0 = 1$).

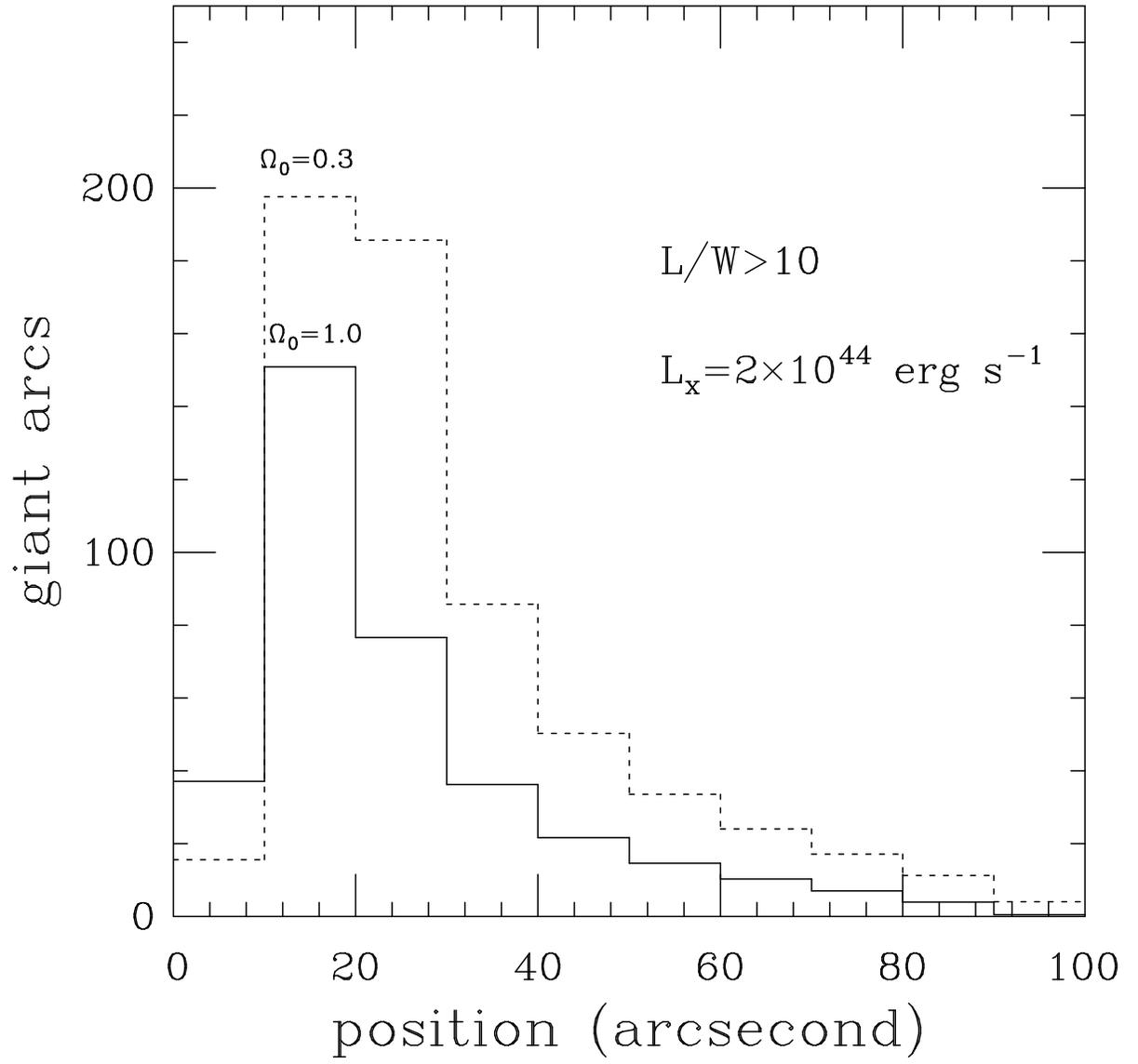

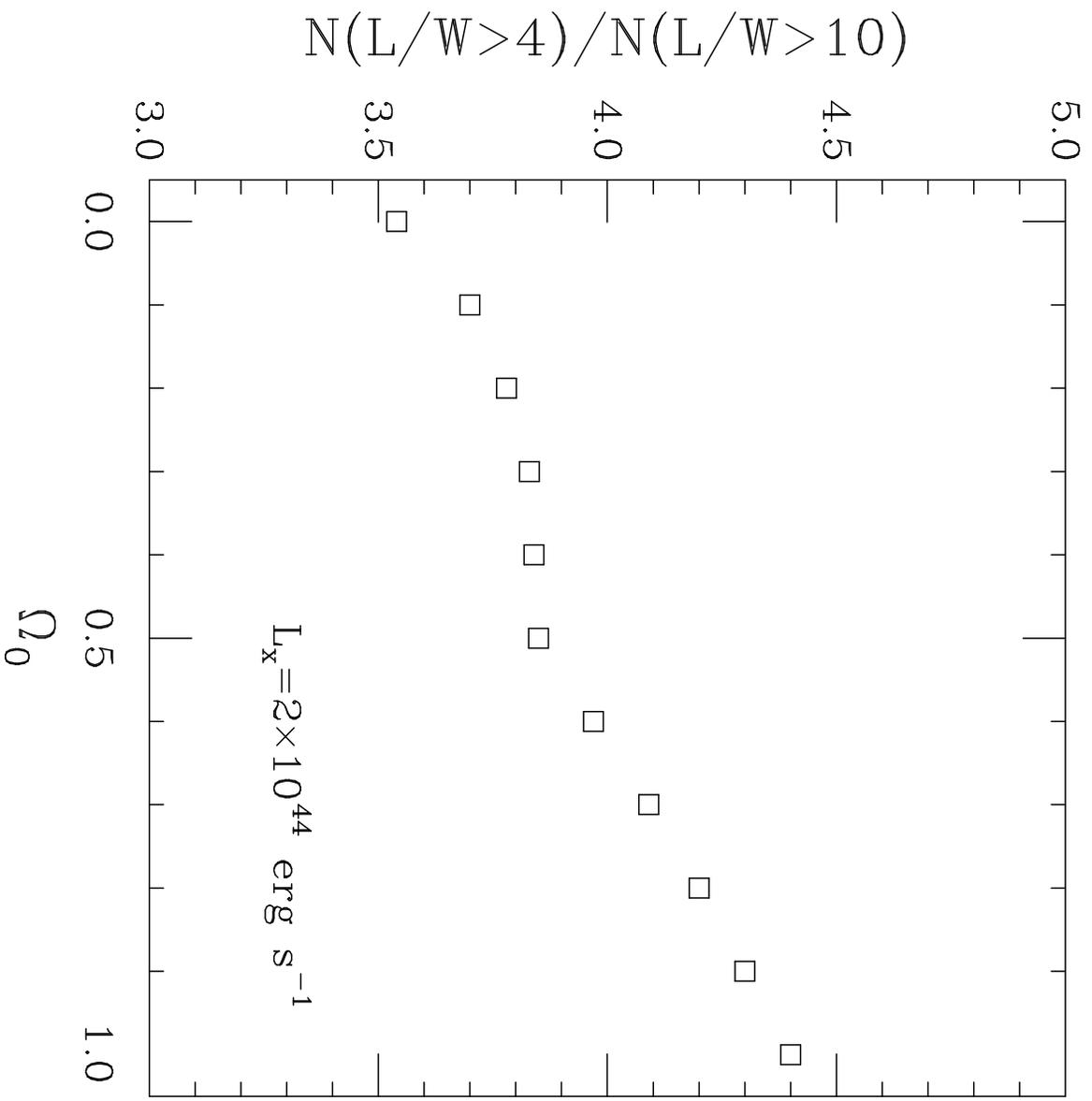